# Quantum Theory of Probability and Decisions

## David Deutsch[1]

*The probabilistic predictions of quantum theory are conventionally obtained from a special probabilistic axiom. But that is unnecessary because all the practical consequences of such predictions follow from the remaining, non-probabilistic, axioms of quantum theory, together with the non-probabilistic part of classical decision theory.*

Introduction

Suppose that a quantum system $\mathfrak{S}$ has density operator $\hat{\rho}$ at the instant when an observable $\hat{X}$ of $\mathfrak{S}$ is to be accurately measured. The outcome of the measurement must be one of the eigenvalues of $\hat{X}$, but quantum theory does not in general specify which. Instead, in conventional formulations of the theory (see e.g. d'Espagnat (1976), Cohen-Tannoudji et al. (1978)), a statement such as the following is taken as axiomatic:

> The probability that the outcome will be $x$ is $\mathbf{Tr}\hat{P}_x\hat{\rho}$, where $\hat{P}_x$ is the projection operator into the space of all eigenvalue-$x$ eigenstates of $\hat{X}$. (1)

---

[1] Centre for Quantum Computation, The Clarendon Laboratory, University of Oxford, Oxford OX1 3PU, UK.

We shall be concerned mainly with measurements of a non-degenerate observable $\hat{X}$ in a pure state $|\psi\rangle$ of $\mathfrak{S}$, in which case the expression for the probability reduces to $|\langle x|\psi\rangle|^2$. I shall in effect prove the probabilistic axiom (1) from the non-probabilistic axioms of quantum theory. The proviso 'in effect' is necessary only because in the conventional formulation the meaning of probability statements is left undefined, while I shall obtain it from the theory. Previous attempts to do this (e.g. Everett (1957), Finkelstein (1963), Hartle (1968), DeWitt and Graham (1973), Ohkuwa (1993)) applied only to infinite sets of measurements (which do not occur in nature), and not to the outcomes of individual measurements (which do). My method is to analyse the behaviour of a rational decision maker who is faced with decisions involving the outcomes of future quantum-mechanical measurements. I shall prove that if he does not assume (1), or any other probabilistic postulate, but does believe the rest of quantum theory, he necessarily makes decisions as if (1) were true. I take the latter to be the effective meaning of (1).

The decision maker is *rational* in the standard decision-theoretic sense (see e.g. Luce and Raiffa (1957)), except that, to avoid circularity, we must omit from the definition of 'rationality' anything that refers directly or indirectly to probabilities. In particular we must not make the standard assumption that a rational decision maker maximises the expectation value of his utility. In this approach, such propositions are to be proved rather than postulated. 'Rationality' in this restricted sense means conformity to a set of constraints on a decision maker's preferences. For example, his preferences must be *transitive:* if he prefers A to B, and B to C, then he must also prefer A to C. Transitive preferences can be summarised by assigning a real number – a utility or *value* – to each possible outcome, in such a way that the decision maker prefers higher-valued outcomes to lower-valued ones.

In classical physics, in situations of perfect knowledge (where one knows all the variables that can affect the outcome, and can calculate how they affect it), the behaviour of a rational decision maker is, up to degeneracy, fully determined by his preferences among the possible outcomes. He chooses one of the options which, he calculates, will cause the highest-valued outcome. In quantum physics, there may not be any 'option which causes the highest-valued outcome', because choosing a given option will in general make possible a range of outcomes, and not even perfect knowledge of the circumstances will allow one to predict which of those one will observe. That is where probability is conventionally introduced, in the form of a probabilistic physical axiom such as (1). I shall show that this is unnecessary.

My assumption that the non-probabilistic part of standard decision theory is applicable in a quantum world is a substantive one. It is not self-evident that rational decision making does not have a radically different character, or that rationality is possible at all, in the presence of quantum-mechanical processes – or, for that matter, in the presence of electromagnetic or any other processes. Nor could any analysis from within physics ever decide what is rational, or what is within the scope of reason. But that is not what I am about here. My objective is to prove something that is conventionally taken as axiomatic (the probabilistic axiom of quantum theory) from other things that are conventionally taken as axiomatic *but do not refer to probability*, namely quantum theory and decision theory, both stripped of their probabilistic axioms.

Deriving a 'tends to' from a 'does'

In cases where quantum theory predicts that a measurement will have a particular outcome, that outcome can, in the conventional formulation, be said to 'have probability 1'. Our decision maker would not put it that way because he does not know what 'probability' means. He would simply predict that that outcome 'will happen' because he knows, from the non-probabilistic part of quantum theory, that if $\hat{X}$ is measured in any of its eigenstates $|x\rangle$, the outcome will be the corresponding eigenvalue *x*. From this he can make a further non-probabilistic prediction, namely that if $\hat{X}$ is measured when $\mathfrak{S}$ is in an arbitrary pure state $|\psi\rangle$, the outcome *x* will be in the set $\{x | \langle x|\psi\rangle \neq 0\}$. To prove this, suppose that the decision maker has made an accurate, non-perturbing measurement of $\hat{X}$ using an apparatus $\mathfrak{A}$, and that he then measures whether the outcome was indeed in the set $\{x | \langle x|\psi\rangle \neq 0\}$. He can do this by measuring the observable

$$\hat{P} = \sum_{\{x | \langle x|\psi\rangle \neq 0\}} |A(x)\rangle\langle A(x)|, \qquad (2)$$

on $\mathfrak{A}$, where $|A(x)\rangle$ is the state in which the apparatus has recorded an outcome *x* for the first measurement. He does not need any probabilistic axiom to predict the outcome of this second measurement: it must be 1, because $\mathfrak{A}$ is in an eigenstate of $\hat{P}$ with eigenvalue 1.

In themselves, predictions of this type are of little practical use, since to identify the set $\{x | \langle x|\psi\rangle \neq 0\}$ one needs to know the state $|\psi\rangle$ with infinite accuracy, which is presumably impossible. Moreover, in the expansions of realistic states there are vast numbers of very small non-vanishing components $\langle x|\psi\rangle|x\rangle$, so identifying the set of *possible* outcomes is not usually very informative. That is why we need what a probabilistic axiom such as (1) conventionally provides, namely a rationale for practical prediction, expectation and decision making in the case of general $|\psi\rangle$ (or $\hat{\rho}$). We need, for instance, to show that if one of the quantities $|\langle x|\psi\rangle|^2$ is overwhelmingly larger than the sum of the others, then it is safe to rely on the corresponding eigenvalue *x* being the outcome of the measurement even though it is *possible* that it will not be; and that if $|\langle x_1|\psi\rangle|^2$ and $|\langle x_2|\psi\rangle|^2$ are equal, it is fair to bet at equal odds on the outcome being $x_1$ or $x_2$; and so on. But we need to show all this without assuming any probabilistic axiom.

Let our decision maker become a player in a simple game in which he knows in advance that $\mathfrak{S}$ is to be prepared in a given pure state $|\psi\rangle$, that an observable $\hat{X}$ of $\mathfrak{S}$ is to be measured, and that he will receive a payoff that depends only on the outcome of the measurement. For convenience, let us consider games in which the measured value of $\hat{X}$ is numerically equal to the utility of the payoff, measured on some suitable utility scale. And let us consider only players for whom the utilities of the possible payoffs can be assigned so as to have an additivity property, namely that the player is indifferent between receiving two separate payoffs with utilities $x_1$ and $x_2$, and receiving a single payoff with utility $x_1+x_2$. Heuristically we may think of $\$$ as a randomising device that displays how much money the player is to be paid, but this is only an approximation because amounts of money will not in general satisfy the additivity condition strictly. The amount of money required to give the player a particular utility $x$ will be a non-linear function of $x$, and will also vary according to the costs and rewards associated with receiving the money under different circumstances.

The *value of a game* to the player is defined as the utility of a hypothetical payoff such that the player is indifferent between playing the game and receiving that payoff unconditionally. Heuristically it is the least upper bound on amounts of money that the player would be willing to pay for the privilege of playing the game. One of the non-probabilistic axioms of decision theory, the *principle of substitutibility*, constrains the values of composite games (games that involve the playing of sub-games). It says that if any of the sub-games is replaced by a game of equal value, then to a rational player, the value of the composite game is unchanged. This really means that if the value of a sub-game depends on the circumstances under which it is played, then there is a way of reinterpreting those circumstances as additional payoffs or conditions of the game, in such a way that the principle of substitutibility will hold – at least for a class of games including those we are considering. Like all the decision-theoretic principles we are applying, this is a substantive assumption, but it is not a *probabilistic* assumption.





The only other axiom we shall need from decision theory concerns *two-player, zero-sum games*. Heuristically, these are games in which the only payoffs consist of money changing hands between two players. In our case they are games in which there are two possible roles – say, A and B – for a player, with the following property: whenever a player would receive a payoff $x$ if he were playing in role A, he would receive $-x$ if he were playing in role B. It is a theorem of classical decision theory that if $\mathcal{V}_A$ and $\mathcal{V}_B$ are the respective values of playing such a game in roles A and B,

$$\mathcal{V}_A + \mathcal{V}_B = 0. \tag{3}$$

In *classical, probabilistic* decision theory, (3) follows from the fact that if the sum of a set of stochastic variables is zero, then the sum of their expectation values is also zero. In *classical, non-probabilistic* decision theory it is trivially true, since each game can have only one payoff, which is equal to the value of the game. Now, since (3) is merely a constraint on the values of various games, and does not relate those values to any probabilities, we may take it as an axiom of our stripped-down, non-probabilistic version of decision theory. Let us call it the *zero-sum rule*.

That games involving quantum measurements *have* values in the above sense is not an independent assumption, but follows from the assumptions I have already made: On being offered the opportunity to play such a game at a given price, knowing $|\psi\rangle$, our player will respond somehow: he will either accept or refuse. His acceptance or refusal will follow a strategy which, given that he is rational, must be expressible in terms of transitive preferences and therefore in terms of a value $\mathcal{V}[|\psi\rangle]$ for each possible game. $\mathcal{V}[|\psi\rangle]$ is the value of playing our game with $\mathfrak{S}$ in the state $|\psi\rangle$ – so a rational decision maker who is given the choice between playing our game with $\mathfrak{S}$ in the state $|\psi_1\rangle$ or $|\psi_2\rangle$, where $\mathcal{V}[|\psi_1\rangle] > \mathcal{V}[|\psi_2\rangle]$, will invariably choose $|\psi_1\rangle$.

I shall prove that





$$\mathcal{V}[|\psi\rangle] = \langle\psi|\hat{X}|\psi\rangle, \qquad (4)$$

and since

$$\langle\psi|\hat{X}|\psi\rangle = \sum_a |\langle x_a|\psi\rangle|^2 x_a, \qquad (5)$$

where the sum is over a complete set $|x_a\rangle$ of eigenstates of $\hat{X}$, it will follow that in making decisions about the outcomes of measurements, a rational decision maker behaves as if he believed that each possible outcome $x_a$ had a probability, given by the conventional formula $|\langle x_a|\psi\rangle|^2$, and as if he were maximising the probabilistic expectation value of the payoff.

This result, which as I said, I take to be the meaning of (1), will also justify what is strictly speaking only an assumption at this stage, namely that the laws of quantum mechanics are consistent with the existence of decision makers whose preferences have the attributes stated above.

I have already noted that in cases where $|\psi\rangle$ is any eigenstate $|x\rangle$ of $\hat{X}$, no probabilistic assumption is needed to predict the outcome of a measurement of $\hat{X}$: it must be $x$. Hence

$$\mathcal{V}[|x\rangle] = x, \qquad (6)$$

which is indeed a special case of (4).

The next simplest of these games are those in which $|\psi\rangle$ is an equal-amplitude superposition of two eigenstates of $\hat{X}$:

$$|\psi\rangle = \frac{1}{\sqrt{2}}(|x_1\rangle + |x_2\rangle). \qquad (7)$$

To conform to (4), we have to show that the value of such games is $\frac{1}{2}(x_1 + x_2)$.





Note that

$$\mathcal{V}\left[\sum_a \lambda_a |x_a + k\rangle\right] = k + \mathcal{V}\left[\sum_a \lambda_a |x_a\rangle\right], \tag{8}$$

where the sums are over all eigenvalues of $\hat{X}$, the $\{\lambda_a\}$ are arbitrary amplitudes and $k$ is an arbitrary constant. This is proved by appealing twice to the additivity of utilities. First, receiving any of the possible payoffs $x_a + k$ of the game referred to on the left of (8) has, by additivity, the same utility as receiving the payoff $x_a$ followed by the payoff $k$. But that sequence of events is physically identical to (and so must have the same utility as) playing the game with $ in the state referred to on the right of (8) and then receiving a payoff $k$, so by additivity again, (8) holds.

It follows from the zero-sum rule (3) that the value of the game of acting as 'banker' in one of these games (i.e. receiving a payoff $-x_a$ when the outcome of the measurement is $x_a$) is the negative of the value of the original game. In other words

$$\mathcal{V}\left[\sum_a \lambda_a |x_a\rangle\right] + \mathcal{V}\left[\sum_a \lambda_a |-x_a\rangle\right] = 0. \tag{9}$$

From (8), with $k = -x_1 - x_2$, and (9), we have

$$-\mathcal{V}\left[\tfrac{1}{\sqrt{2}}(|x_2\rangle + |x_1\rangle)\right] = -x_1 - x_2 + \mathcal{V}\left[\tfrac{1}{\sqrt{2}}(|x_1\rangle + |x_2\rangle)\right], \tag{10}$$

i.e.

$$\mathcal{V}\left[\tfrac{1}{\sqrt{2}}(|x_1\rangle + |x_2\rangle)\right] = \tfrac{1}{2}(x_1 + x_2), \tag{11}$$

as required.

In (11) we have determined the value of a game whose outcome is indeterminate. That is the pivotal result of this paper. We shall see that the general result (4), and therefore in effect the axiom (1), follow quite straightforwardly from it. Yet we have derived it strictly from non-probabilistic quantum theory and non-probabilistic





decision theory, starting with valuations $x_1$ and $x_2$ of determinate games. Thus we see that quantum theory permits what philosophy would hitherto have regarded as a formal impossibility, akin to 'deriving an ought from an is', namely deriving a probability statement from a factual statement. This could be called deriving a 'tends to' from a 'does'.

The general case

The next step towards proving (4) is to generalise (11) to equal-amplitude superpositions of *n* eigenstates of $\hat{X}$:

$$\mathcal{V}\left[\frac{1}{\sqrt{n}}\left(|x_1\rangle+|x_2\rangle+\ldots+|x_n\rangle\right)\right]=\frac{1}{n}(x_1+x_2+\ldots+x_n). \tag{12}$$

The proof of (12) is by induction, in two stages. The first, covering only the cases where $n=2^m$ for some integer *m*, is on ascending values of *m*, and the second is on descending values of *n* to cover the remaining cases. The first follows immediately from the principle of substitutibility on considering a game with two equal-amplitude outcomes, each of which is replaced by a game with $2^{m-1}$ equal-amplitude outcomes. For the second, note that if every possible outcome of a game has utility $v$ then the game itself has value $v$. This is a consequence of additivity, since playing such a game amounts to performing a measurement whose outcome is ignored, and then receiving a payoff $v$ unconditionally. Hence, if $|\psi_1\rangle$ and $|\psi_2\rangle$ are superpositions of eigenstates of $\hat{X}$ chosen respectively from two non-intersecting sets of eigenstates, and α and β are complex numbers, and $\mathcal{V}[|\psi_1\rangle]=\mathcal{V}[|\psi_2\rangle]=v$, substitutibility implies that

$$\mathcal{V}\left[\frac{\alpha|\psi_1\rangle+\beta|\psi_2\rangle}{\sqrt{|\alpha|^2+|\beta|^2}}\right]=v. \tag{13}$$

Setting





$$\left. \begin{array}{cc} |\psi_1\rangle = \dfrac{1}{\sqrt{n-1}}(|x_1\rangle + |x_2\rangle + \ldots + |x_{n-1}\rangle), & |\psi_2\rangle = \big|\mathcal{V}\big[|\psi_1\rangle\big]\big\rangle, \\ \alpha = \sqrt{n-1}, & \beta = 1, \end{array} \right\} \quad (14)$$

in (13), and assuming (12) as the inductive hypothesis, we have

$$\frac{1}{n}\left(x_1 + x_2 + \ldots + x_{n-1} + \mathcal{V}\big[|\psi_1\rangle\big]\right) = \mathcal{V}\big[|\psi_1\rangle\big], \quad (15)$$

which is equivalent to (12) with $n-1$ replacing $n$. The substitution (14) is valid only if $\mathcal{V}\big[|\psi_1\rangle\big]$ is different from each of the eigenvalues $x_1 \ldots x_{n-1}$, but this can always be arranged by choosing a suitable eigenvalue to label as '$x_n$' in (12).

Now we can generalise to a case with unequal amplitudes by showing that

$$\mathcal{V}\left[\sqrt{\frac{m}{n}}|x_1\rangle + \sqrt{\frac{n-m}{n}}|x_2\rangle\right] = \frac{mx_1 + (n-m)x_2}{n}, \quad (16)$$

where $m$ and $n$ are integers. One way of measuring $\hat{X}$ when playing the game referred to in (16) is to place an auxiliary system $\mathfrak{T}$ in one of the two states

$$\frac{1}{\sqrt{m}}\sum_{a=1}^{m}|y_a\rangle \quad \text{or} \quad \frac{1}{\sqrt{n-m}}\sum_{a=m+1}^{n}|y_a\rangle, \quad (17)$$

according as $\hat{X}$ takes the value $x_1$ or $x_2$ respectively, where the expansions are in terms of eigenstates of an observable $\hat{Y}$ of $\mathfrak{T}$, and the $n$ eigenvalues $\{y_a\}$ are all distinct. If the operation on $\mathfrak{T}$ is performed coherently, the joint state of $\mathfrak{S}$ and $\mathfrak{T}$ becomes

$$\frac{1}{\sqrt{n}}\left(\sum_{a=1}^{m}|x_1\rangle|y_a\rangle + \sum_{a=m+1}^{n}|x_2\rangle|y_a\rangle\right). \quad (18)$$

Then we measure $\hat{Y}$. If the outcome is one of the $\{y_a\}$ for $1 \leq a \leq m$, we have measured $\hat{X}$ to have the value $x_1$. Otherwise we have measured it to have the value $x_2$.

Let the $\{y_a\}$ be chosen to have the additional properties that





$$\sum_{a=1}^{m} y_a = \sum_{a=m+1}^{n} y_a = 0, \tag{19}$$

and that the *n* values $\{x_1 + y_a \, (1 \le a \le m)\}$, $\{x_2 + y_a \, (m < a \le n)\}$ are all distinct. Then the player will be indifferent to playing a further game in which he receives the measured value of $\hat{Y}$, for (19) ensures that both versions of that game (played after the first payoff was respectively $x_1$ or $x_2$) have value zero. In other words, the composite game played with $\hat{X}$ and $\hat{Y}$ consecutively has the same value as the game played with $\hat{X}$ alone. However, because of additivity the composite game also has the same value as the game in which a single measurement is made of the observable $\hat{X} \otimes \hat{1} + \hat{1} \otimes \hat{Y}$, and in terms of eigenstates of that observable, the state (18) is an equal-amplitude superposition

$$\frac{1}{\sqrt{n}} \left( \sum_{a=1}^{m} |x_1 + y_a\rangle + \sum_{a=m+1}^{n} |x_2 + y_a\rangle \right). \tag{20}$$

So according to (12) and (19) the value of the game is

$$\frac{1}{n} \left[ \sum_{a=1}^{m} (x_1 + y_a) + \sum_{a=m+1}^{n} (x_2 + y_a) \right] = \frac{m x_1 + (n - m) x_2}{n}, \tag{21}$$

as required. By replacing the payoffs $x_1$ or $x_2$ by games with those values, one can obtain an analogous result for any finite superposition

$$\sum_{a} \sqrt{p_a} |x_a\rangle \qquad \left( \sum_{a} p_a = 1 \right) \tag{22}$$

whose coefficients $\{\sqrt{p_a}\}$ are non-negative square roots of rational numbers.

To remove the restriction that the $\{p_a\}$ be rational, consider yet another class of games. In these, $\mathfrak{S}$ undergoes some unitary evolution U after it is prepared in its initial state $|\psi\rangle$ but before $\hat{X}$ is measured. In other words the state evolves to $U|\psi\rangle$, so the value of the transformed game is

$$\mathcal{V}_U[|\psi\rangle] = \mathcal{V}[U|\psi\rangle]. \tag{23}$$





Now, if U transforms each eigenstate $|x_a\rangle$ of $\hat{X}$ appearing in the expansion of $|\psi\rangle$ to an eigenstate $|x_{a'}\rangle$ with higher eigenvalue, then by additivity the value of the transformed game exceeds that of the original game. The same is true if each $|x_a\rangle$ evolves into a *superposition* of higher-eigenvalue eigenstates, by additivity and substitutibility, for the player will agree to a transformation that is guaranteed to increase his payoff, albeit by an unknown amount.

Consider transformations U that evolve each eigenstate of $\hat{X}$ that appears in the expansion of $|\psi\rangle$ into a superposition of itself and higher-eigenvalue eigenstates. Because of the denseness of the rational numbers in the reals, there exist arbitrarily slight transformations which have that property and evolve $|\psi\rangle$ into a form $U|\psi\rangle$ such that the squares of all the coefficients in the expansion of $U|\psi\rangle$ in eigenstates of $\hat{X}$ are rational. Each game played with such a state $U|\psi\rangle$ is at least as valuable as the original game, and the values of such games have a lower bound

$$\sum_a |\langle x_a|\psi\rangle|^2 x_a . \qquad (24)$$

Similarly, the values of transformed games where the squares of the coefficients in the expansion of $U|\psi\rangle$ are rational, and where the games are at *most* as valuable as the original game, have (24) as their *upper* bound. It follows that (24) is the value of the original game, as required.

To prove that (24) remains the value of the game if the coefficients $\langle x_a|\psi\rangle$ are arbitrary complex amplitudes, let $\{\phi_a\}$ be a set of arbitrary phases. Since the unitary evolution defined by

$$|x_a\rangle \to e^{i\phi_a}|x_a\rangle, \qquad (25)$$

performed after the measurement of $\hat{X}$, does not alter the payoff, the player is indifferent as to whether it occurs or not. But the final state following such evolution is the same as it would have been if (25) had occurred before the measurement of $\hat{X}$. Consequently, by additivity and substitutibility, state transformations of the form (25) do not affect the values of our games. That completes the proof of (4) for general pure states $|\psi\rangle$ of systems with finite-dimensional state spaces.

Generalising these results to cases where $\mathfrak{S}$ is not in a pure state is trivial if $\mathfrak{S}$ is part of a larger system that is in a pure state, for then every measurement on $\mathfrak{S}$ is also a measurement on the larger system. Further generalisation to exotic situations in which the universe as a whole may be in a mixed state (Hawking (1976), Deutsch (1991)), is left as an exercise for the reader.

Conclusions

No probabilistic axiom is required in quantum theory. A decision maker who believes only the non-probabilistic part of the theory, and is 'rational' in the sense defined by a strictly non-probabilistic restriction of classical decision theory, will make all decisions that depend on predicting the outcomes of measurements as if those outcomes were determined by stochastic processes, with probabilities given by axiom (1). (However, in other respects he will not behave as if he believed that stochastic processes occur. For instance if *asked* whether they occur he will certainly reply 'no', because the non-probabilistic axioms of quantum theory require the state to evolve in a continuous and deterministic way.)

Likewise, no probabilistic axiom is required in decision theory. Classical stochastic processes do not occur in nature, so decision theory need not concern itself with them. Where probabilities arise from quantum indeterminacy, we have proved that a rational decision maker will maximise the expectation value of his utility. Where numbers obeying the probability calculus arise in any other context, regarding them as probabilities in the decision-theoretic sense needs to be independently justified.

The usual probabilistic terminology of quantum theory is justifiable in the light of the results of this paper, provided that one understands it all as referring ultimately to the behaviour of rational decision makers. For instance, defining the *expectation value* of an outcome in such terms does relate it to the common-sense notion of 'expectation'. For suppose that a rational decision maker does indeed pay an amount $\mathcal{V}[|\psi\rangle]$ for the privilege of playing a game of that value. He is, prospectively, indifferent to doing so. But subsequently, he will have received a payoff which may or may not be $\mathcal{V}[|\psi\rangle]$. If it is not, then although he knew in advance that this could happen, he will no longer be indifferent, for he will have made either a definite profit or a definite loss on the transaction. This deviation from indifference is brought about by his discovering that the actual payoff was respectively more or less than his prior valuation. So $\mathcal{V}[|\psi\rangle]$ is indeed the payoff that a rational player 'expects' to receive, in the sense that it is the one from which he benefits neither more nor less than he has budgeted for. Of course in another sense he may well not be expecting that outcome: for instance when $\mathcal{V}[|\psi\rangle]$ is not an eigenvalue of $\hat{X}$, he knows that $\mathcal{V}[|\psi\rangle]$ is not one of the possible payoffs of the game.

Similarly, an outcome may be said to be *random* if it is unpredictable, and if (as will always be the case in quantum theory), enough information to calculate its expectation value exists somewhere. The term *probability* itself can be defined in this way, working backwards from expectation values, and that is the sense in which (1) follows from non-probabilistic postulates. Thus predictions such as 'the probability of outcome $x$ is $|\langle x|\psi\rangle|^2$' become implications of a purely factual theory, rather than axioms whose physical meanings are undefined.

**Acknowledgements**

I am grateful to Prof. B.S. DeWitt for drawing my attention to the problem addressed in this paper. I also wish to thank him and Dr M.J. Lockwood and Prof. F.J. Tipler for stimulating conversations on this subject, and Prof. D.N. Page for pointing out some significant errors in previous drafts.